# Effects of structure and temperature on the nature of excitons in the Mo$_{0.6}$W$_{0.4}$S$_2$ alloy

**AUTHORS**

Deepika Poonia,[1,†] Nisha Singh,[1,†] Jeff J.P.M. Schulpen,[‡] Marco van der Laan,[¶] Sourav Maiti,[†] Michele Failla,[†] Sachin Kinge,[§] Ageeth A. Bol,[‡] Peter Schall,[¶] and Laurens D.A. Siebbeles*[†]

**AFFILIATIONS**

[†]*Optoelectronic Materials Section, Department of Chemical Engineering, Delft University of Technology, 2629 HZ Delft, The Netherlands*

[‡] *Department of Applied Physics, Eindhoven University of Technology, P.O. Box 513, 5600 MB Eindhoven, The Netherlands*

[¶] *Institute of Physics, University of Amsterdam, 1098 XH Amsterdam, The Netherlands*

[§]*Materials Research & Development, Toyota Motor Europe, B1930 Zaventem, Belgium*

[1]Both authors contributed equally to this work.

**ABSTRACT**

We have studied the nature of excitons in the transition metal dichalcogenide alloy $Mo_{0.6}W_{0.4}S_2$, compared to pure $MoS_2$ and $WS_2$ grown by atomic layer deposition (ALD). For this, optical absorption/transmission spectroscopy and time-dependent density functional theory (TDDFT) were used. Effects of temperature on the A and B exciton peak energies and linewidths in the optical transmission spectra were compared between the alloy and pure $MoS_2$ and $WS_2$. On increasing the temperature from 25 K to 293 K the energy of the A and B exciton peaks decreases, while their linewidth increases due to exciton-phonon interactions. The exciton-phonon interactions in the alloy are closer to those for $MoS_2$ than $WS_2$. This suggests that the exciton wave functions in the alloy have a larger amplitude on Mo atoms than on W atoms. The experimental absorption spectra could be reproduced by TDDFT calculations. Interestingly, for the alloy the Mo and W atoms had to be distributed over all layers. Conversely, we could not reproduce the experimental alloy spectrum by calculations on a structure with alternating layers, in which every other layer contains only Mo atoms and the layers in between also W atoms. For the latter atomic arrangement, the TDDFT calculations yielded an additional optical absorption peak that could be due to excitons with some charge transfer character. From these results we conclude that ALD yields an alloy in which Mo and W atoms are distributed uniformly among all layers.

**TOC Graphics**

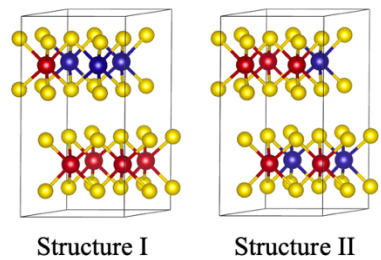 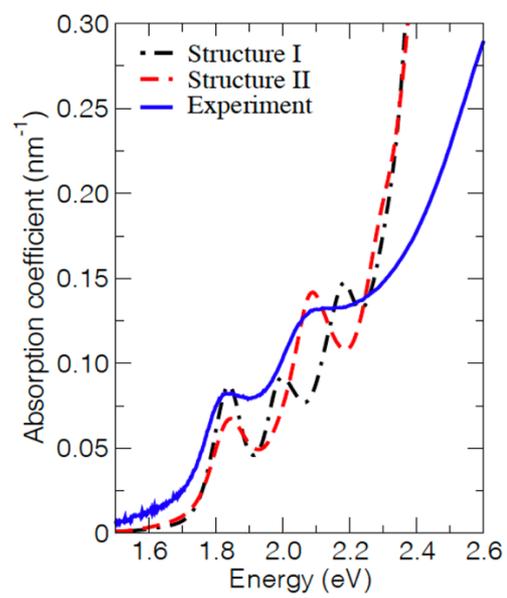

1. **INTRODUCTION**

Layered van der Waals materials, in particular transition metal dichalcogenides (TMDCs), have gained considerable interest due to prospects for applications in e.g., photodetectors,[1,2] sensors,[3,4] and solar cells.[5–7] These materials consist of layers in which transition metal atoms are covalently bound to chalcogen (S, Se, Te) atoms. The layers are stacked on top of each other and held together by van der Waals forces.[8,9] TMDCs with chemical composition $MX_2$ (M= Mo, W, etc. and X = S, Se) have been studied extensively owing to their direct bandgap in monolayers,[10] valley selective optical coupling,[11] and large exciton binding energies.[12] The electronic and optical properties of these TMDCs can be tuned by alloying. Alloying has been used to vary the relative content (y) of the transition metal or chalcogen atoms and obtain layers of $M_y M'_{1-y} X_2$ or $MX_{2y} X'_{2(1-y)}$.[13,14]

For optoelectronic applications, understanding of electron-phonon and exciton-phonon interactions is important. The strength of these interactions governs charge transport,[15] bandgap renormalization,[16] optical heating of the lattice[17], and intervalley scattering of excitons.[18–20] In this regard, effects of temperature on optical absorption and photoluminescence spectra can provide information about the coupling strength between excitons and phonons in TMDCs.[18–21]

We investigated effects of temperature on exciton peak energies and linewidths in the optical transmission spectra of the $Mo_{0.6}W_{0.4}S_2$ alloy and pure $MoS_2$ and $WS_2$. The samples were grown by atomic layer deposition (ALD).[22] The almost equal content of Mo and W atoms in the alloy is of interest since it offers the possibility to realize intimate mixing of the transition metal atoms, rather than having separate domains consisting of one atom type only. To elucidate effects of the relative arrangement of Mo and W atoms in the alloy, we compared the measured spectra with results from *ab-initio* time-dependent density functional theory (TDDFT) calculations. For this purpose, we constructed supercells having different positions of the metal atoms in the crystal structure of the alloy. The TDDFT calculations reproduced the experimental spectrum of the alloy for structures in which all layers contain both Mo and W atoms. In contrast, calculations on a structure containing W atoms in individual layers that are separated by layers containing only Mo atoms do not reproduce the experimental spectrum. From the latter, we infer that the ALD growth yields structures with a predominantly homogeneous spatial distribution of Mo and W atoms.

## 2. METHODS

**2.1. Temperature-dependent optical transmission measurements**. We used our previously reported ALD procedure to grow thin films of $MoS_2$, $WS_2$ and the $Mo_{0.6}W_{0.4}S_2$ alloy, with thickness of 6.3 nm, 4.1 nm and 5.2 nm, respectively, on quartz substrates.[22] The alloy was grown using an ALD supercycle length of 2 cycles (consisting of 1 $MoS_2$ cycle and 1 $WS_2$ cycle), to realize fine mixing of the Mo and W atoms. The composition was determined by XPS.[22] The separation between adjacent layers in these materials is ~0.6 nm, so the film thicknesses correspond to 10-11, 6-7 and 8-9 layers, respectively.

The optical transmission of the samples was measured using a home-built setup containing a DH-2000 halogen light source and an Ocean-optics Maya 2000 spectrometer. To vary the temperature, the samples were placed under vacuum in a He-closed cycle cryostat. These measurements yield the fraction of light transmitted, *T*, through the sample as a function of photon energy and temperature.

For comparison of the optical properties of the samples with the optical absorption coefficient from TDDFT calculations (see Section 3.3), we determined the optical density, $OD$, using a Perkin Elmer Lambda 1050 spectrometer with an integrating sphere. This could be done only at room temperature, since the spectrometer is not equipped with a cryostat. Placing the sample in front of the light entrance of the integrating sphere yields $T$, and placing it in the center provides $T+R$, where $R$ is the fraction of light reflected. Results of $1-T$, $R$, and the fraction of light absorbed $A = 1 - R - T$ are shown in Figure S1 for the pure compounds and the alloy. The optical density is obtained using the relation $OD = -\log\left(\frac{T}{1-R}\right)$. The optical absorption coefficient, $\alpha$, of a film with thickness $L$ is related to the $OD$ according to $e^{-\alpha L} = 10^{-OD}$, giving $\alpha = OD \ln(10)/L$.

**2.2. TDDFT calculations of optical absorption coefficients.** Electronic structure calculations were performed using the all-electron full-potential linearized augmented plane wave (LAPW) code Elk[23] with PBE (GGA) functionals.[24] For all materials a hexagonal crystal structure (2H) was used, with the experimental lattice constants of 3.169 Å and 12.324 Å for MoS$_2$[25] and 3.153 Å and 12.323 Å for WS$_2$.[26] A 2×2×1 supercell is constructed to study the Mo$_{0.625}$W$_{0.375}$S$_2$ alloy with the lattice parameters 6.338 Å and 12.324 Å obtained by doubling the MoS$_2$ unit cell. Calculation of the dielectric response functions from TDDFT requires a dense k-point grid to sample the Brillouin zone (BZ); hence a k-point grid of 16×16×8 for the primitive unit cell and a 8×8×8 k-point grid for the supercell are used. The set of LAPW basis functions is defined by specifying a cut off parameter $|\mathbf{k} + \mathbf{G}|_{max}$ whose value is set to 7.0 Bohr$^{-1}$. Additionally, the response is calculated using G vectors of length 1.5 Bohr$^{-1}$. The number of conduction bands included in the calculations is 24 for both MoS$_2$, WS$_2$, and 96 for the alloy.

In TDDFT, a Dyson-like equation is solved to obtain the dielectric response function[27] whose real and imaginary parts can be used to obtain the optical absorption coefficient $\alpha$.[28] The method to obtain optical response functions is a two-step procedure. Firstly, a ground-state calculation is done to obtain the converged density and potentials. Next, the dielectric functions of $MoS_2$, $WS_2$, and the $Mo_{0.625}W_{0.375}S_2$ alloy are calculated as a function of photon energy using the bootstrap kernel,[29] as it is capable of capturing excitons in the TDDFT calculations. The dielectric functions thus obtained were broadened by 80 meV for $MoS_2$, $WS_2$ and 54 meV for the alloy to obtain the best matches with the experimental optical absorption coefficient spectra ($\alpha$). Note that the broadening thus introduced in the calculated spectra does not explain the exciton linewidths in the experimental spectra.

The absolute values of exciton energies with respect to the ground state cannot be accurately captured by TDDFT, due to the well-known bandgap problem. To overcome this, we employed the so-called 'scissor operator' method that shifts the entire optical absorption spectrum ($\alpha$) in energy. To reproduce the lowest experimental exciton energy, we used energy shifts of 0.03 eV, 0.08 eV, and 0.06 eV for $MoS_2$, $WS_2$, and the $Mo_{0.625}W_{0.375}S_2$ alloy, respectively.

## 3. RESULTS AND DISCUSSION

**3.1 Optical transmission spectra.** Figure 1(a) shows optical transmission spectra of $MoS_2$, $WS_2$, and the $Mo_{0.6}W_{0.4}S_2$ alloy at room temperature (293 K). These spectra show the magnitude of 1-*T*, which is the fraction of incident light that is not transmitted through the sample. The spectra of $MoS_2$ and $WS_2$ agree with previous results.[8,30] Two distinct peaks (marked by A and B) can be seen in all three materials. The peaks are due to photoexcitation from the ground state to A and B exciton states. The energies of these peaks are determined by spin-orbit coupling and interlayer interactions at the K and K' point of the Brillouin zone (BZ).[31–33] Towards the higher energy side, a broad absorption feature is observed (often addressed as C exciton), which originates from multiple transitions from the highest valence band to the lowest conduction bands near the Γ point of the BZ.[34] On lowering the temperature to 25 K (Figure 1(b)), the exciton peaks of all three materials become narrower and shift to higher energy.

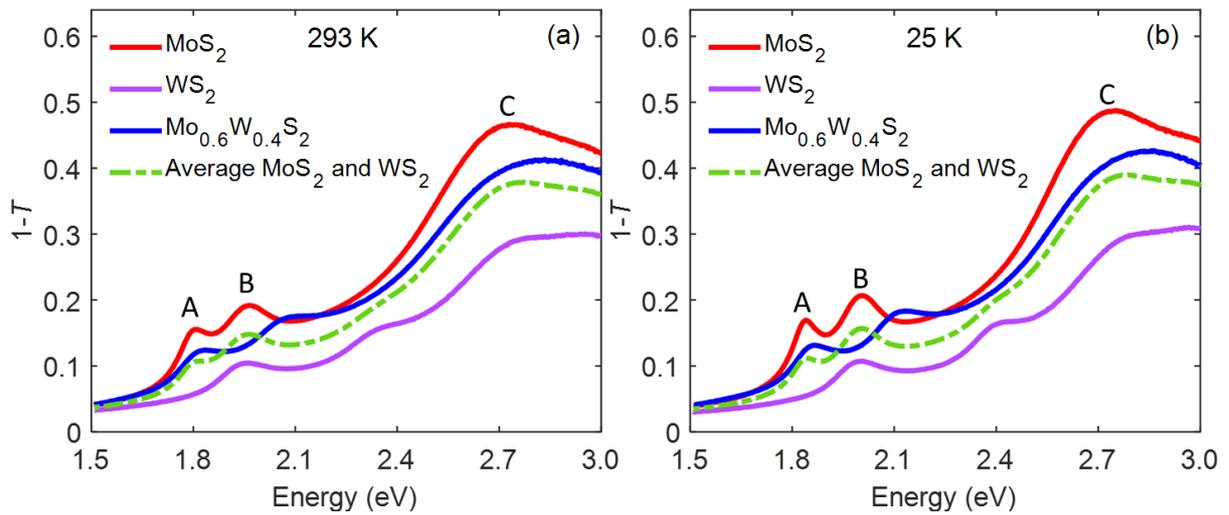

**Figure 1.** (a) Room-temperature (293 K) and (b) low temperature (25 K) optical transmission spectra of $MoS_2$ (red), $WS_2$ (magenta) and the $Mo_{0.6}W_{0.4}S_2$ alloy (blue). The dashed green curves are the average of the $MoS_2$ and $WS_2$ spectra.

To gain qualitative insight into the effect of alloying, we also show the average of the spectra of pure $MoS_2$ and $WS_2$ as green dashed curves in Figure 1 (a quantitative comparison of the measured *OD* and results from TDDFT is discussed in Section 3.3). The average spectra at 293 K and 25 K both differ from the spectra of the alloy. Most strikingly, the B exciton peak of the alloy appears at significantly higher energy than in the average spectra. These differences indicate that formation of excitons in domains consisting of either predominantly MoS$_2$ or WS$_2$ is unlikely. As a consequence, the probability that photoexcitation leads to formation of a charge transfer exciton at a boundary between these material domains is small. Indeed, the peak of the A exciton in the alloy spectrum appears at higher energy than that in the spectrum of MoS$_2$, while that of a charge transfer exciton would be at lower energy.

Inspection of the transmission spectra of the alloy points towards closer similarities to MoS$_2$ than to WS$_2$. Despite intimate mixing and nearly equal Mo and W content in the alloy,[22] the energies of the A and B exciton in the alloy are closer to those of pure MoS$_2$, as also found for monolayers before.[14] This suggests that the wave functions of excitons in the Mo$_{0.6}$W$_{0.4}$S$_2$ alloy have a larger amplitude on Mo atoms than on W atoms. The latter agrees with charge density distributions for the highest occupied and lowest unoccupied orbitals obtained from DFT calculations.[14]

**3.2. Temperature dependence of A and B exciton peak energies and linewidths.** To get further insight into the relative contributions of Mo and W atoms to the character of excitons, we compare the effects of exciton-phonon coupling in the Mo$_{0.6}$W$_{0.4}$S$_2$ alloy with those in MoS$_2$ and WS$_2$. We studied electron-phonon coupling by analysis of the temperature dependence of exciton peak energies and linewidths in the transmission spectra, as outlined in Section 2 of the Supporting Information. The peaks due to A and B excitons could each be described by a Lorentzian function with linewidth $\Gamma_X$, (where X = A, B), which is defined as the full-width at half-maximum (FWHM), see Equation S1. The contribution of optical reflection, below band gap absorption due to defects,[35] and the broad C absorption feature at higher energy in the optical transmission spectra in Figure 1 could be described by two Gaussian functions. The total fit function thus consists of two Lorentzian and two Gaussian functions, see Equation S1. Figure S2 shows that the fits reproduce the experimental transmission spectra very well.

Figures 2 and 3 show the temperature dependence of the A and B exciton peak energies and linewidths, as obtained from fits of Equation S1 to the experimental transmission spectra. At all temperatures, the peak energies and linewidths of the $Mo_{0.6}W_{0.4}S_2$ alloy are closer to those of $MoS_2$ than $WS_2$. This further supports the idea that excitons have more Mo than W character, as we already inferred above from Figure 1.

The decrease of the exciton peak energies with increasing temperature is due to the availability of more phonons at higher temperatures that can be absorbed upon photoexcitation from the electronic ground state to an exciton state, as well as electron-phonon coupling due to interaction between the motion of electrons and atomic nuclei (change of bond lengths and breakdown of the Born-Oppenheimer approximation).[36–38] Following previous studies[21,38–40] we describe the temperature dependence of the exciton peak energies by the semi-empirical O'Donnell equation[41]

$$E_X = E_{0,X} - S_X \langle \hbar\omega_X \rangle \left[ \coth\left(\frac{\langle \hbar\omega_X \rangle}{2 k_B T}\right) - 1 \right] \quad (1)$$

where X = A, B denotes the exciton type and $k_B$ and $\hbar$ are the Boltzmann and the reduced Planck constant. In Equation 1, $E_{0,X}$ is the exciton peak energy at zero temperature, $S_X$ is a dimensionless constant that increases with the exciton-phonon coupling strength and $\langle \hbar\omega_X \rangle$ is the coupling-weighted average of the phonon energies that interact with the exciton.[42]

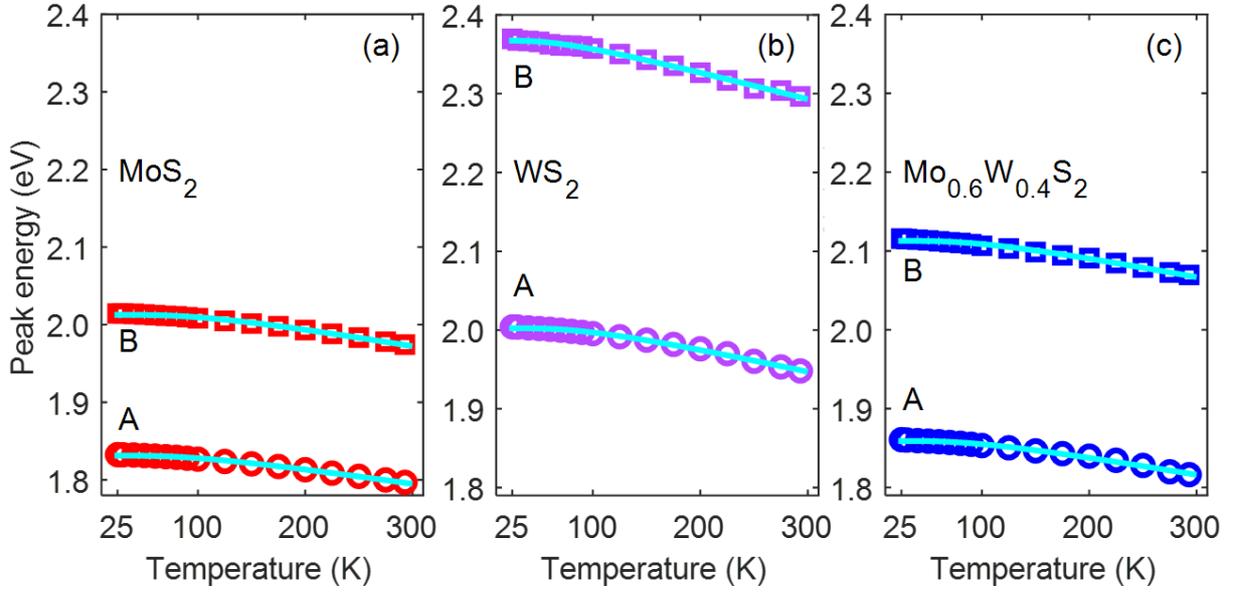

**Figure 2.** Temperature dependence of the A and B exciton peak energies for (a) MoS$_2$, (b) WS$_2$, and (c) the Mo$_{0.6}$W$_{0.4}$S$_2$ alloy, obtained from the measured transmission spectra (markers). The solid cyan curves are fits of Equation 1 to the experimental peak energies.

Fits of Equation 1 to the A and B exciton peak energies with $E_{0,X}$, $S_X$ and $\langle \hbar\omega_X \rangle$ as adjustable parameters are shown as solid cyan curves in Figure 2. Equation 1 reproduces the temperature dependence of the exciton peak energies very well and the values of the fit parameters are presented in Table 1. The exciton peak energies $E_{0,A}$ and $E_{0,B}$ for the alloy are closer to those for MoS$_2$ than for WS$_2$. In addition, the fitted values near 1.5 of $S_A$ and $S_B$ for the alloy are similar to those of MoS$_2$, while they are about 25% smaller than the values near 2.0 obtained for WS$_2$. These findings corroborate our notice in Section 3.1 that exciton wave functions in the alloy have a larger amplitude on Mo atoms than on W atoms, so that the former has a predominant effect on exciton-phonon coupling. To within the experimental uncertainty, the average phonon energies $\langle \hbar\omega_X \rangle$ for both A and B excitons are similar for all three materials and are close to the value of 22.1 meV reported for MoS$_2$ and WS$_2$ in literature.[43,44]

We analyze the temperature dependence of the linewidths of the Lorentzians in Equation S1 of the A and B exciton peaks by using the following expression[45]

$$\Gamma_X = \Gamma_{X,I} + \frac{\Gamma_{X,ph}}{\left(e^{\frac{\langle \hbar\omega_X \rangle}{k_B T}} - 1\right)} \quad (2)$$

The first term at the right-hand side of Equation (2), $\Gamma_{X,I}$, represents inhomogeneous linewidth broadening induced by temperature-independent mechanisms, such as scattering of excitons on structural defects or impurities. The second term describes exciton-phonon scattering for both absorption and emission of phonons. The average energies of phonons that couple with excitons, $\langle \hbar\omega_X \rangle$, were taken equal to the values obtained from fitting Equation 1 to the peak energies, see Table 1.

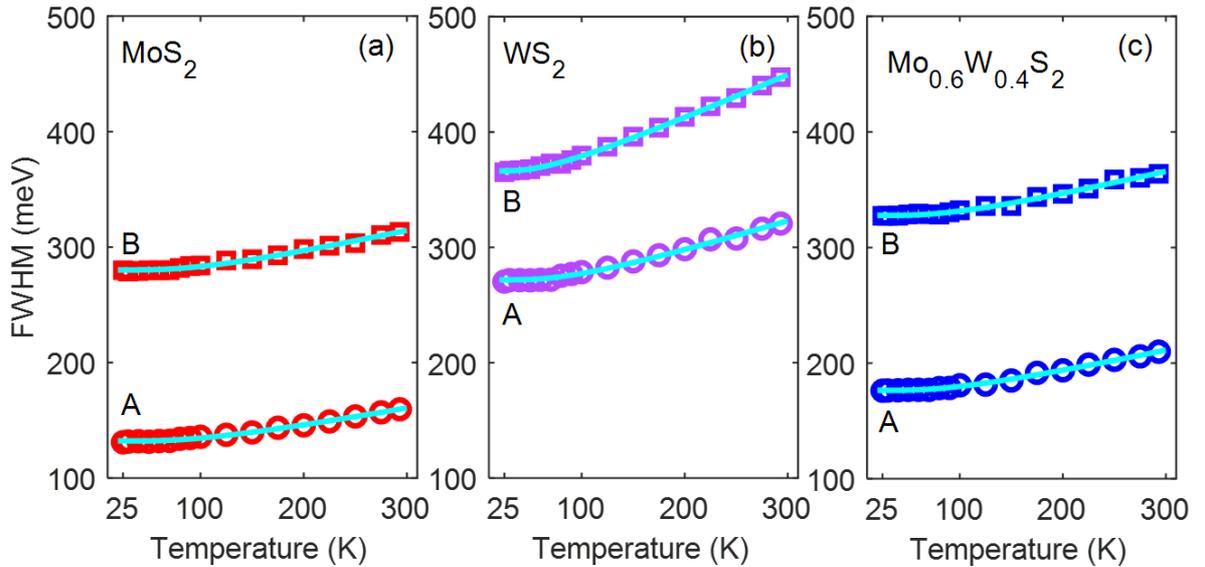

**Figure 3.** Temperature dependence of the linewidths (FWHM, markers) of the A and B exciton peaks for (a) $MoS_2$, (b) $WS_2$ and (c) the $Mo_{0.6}W_{0.4}S_2$ alloy. The solid cyan lines are fits to the experimental data.

The solid cyan lines in Figures 3(a)-(c) are the least-squares fits of Equation 2 to the FWHM values, with the latter obtained from fits of Equation S1 to the optical transmission spectra in Figure S2. The results for the inhomogeneous broadening, $\Gamma_{X,I}$, and the broadening due to exciton-phonon scattering, $\Gamma_{X,ph}$, are presented in Table 1. For each of the three materials, the values of the inhomogeneous broadening of the A exciton, $\Gamma_{A,I}$, are smaller than that of the B exciton, $\Gamma_{B,I}$, similar to results for single crystals.[46] Interestingly, the values of both $\Gamma_{A,ph}$ and $\Gamma_{B,ph}$ of the alloy are close to the corresponding values for $MoS_2$, while they are significantly lower than those for $WS_2$. This is in line with the exciton peak energies and the values of $S_A$ and $S_B$ for the alloy being nearest to those of $MoS_2$, as discussed above. The larger exciton-phonon scattering rate for B excitons can be due to the additional ultrafast decay channel of B excitons involving their relaxation to A excitons by emission of phonons, as discussed previously.[47]

Our values of $E_{0,X}$, $S_X$, $\langle\hbar\omega_X\rangle$ and $\Gamma_{X,ph}$ for ALD grown $MoS_2$ and $WS_2$ films are within the range reported for mono- or few-layer TMDC samples that were obtained by mechanical exfoliation or chemical vapor deposition (CVD),[21,39,40,44,48–50] and CVD grown bulk samples.[44,51] Note that the values of these parameters can vary from one sample to another due to differences in sample preparation, dielectric environment (in particular for mono- and few-layer samples), etc. Our values for the inhomogeneous linewidth broadening, $\Gamma_{X,I}$, are higher than those that Ho et al.[51] obtained from temperature-dependent piezoreflectance measurements on CVD-grown crystals of $MoS_2$, $WS_2$ and $Mo_xW_{1-x}S_2$ alloys. This may result from a larger degree of structural disorder in our ALD-grown samples. Indeed the grain size in ALD grown samples is ~10 nm, which is much smaller than for CVD-grown crystals.[52] Interestingly, the values of the electron-LO phonon coupling strength, $\Gamma_{X,ph}$, reported by Ho et al.[51] are a factor of 2-3 higher than ours. This could be due to the fact that their piezoreflectance measurements probe excitons near the sample surface, which would then appear to couple to surface-phonons with higher strength than the bulk exciton-phonon coupling probed in our experiments.

**Table 1.** Fitted values of the exciton-phonon coupling strength, $S_X$, the average phonon energy, $\langle \hbar\omega_X \rangle$, inhomogeneous linewidth broadening, $\Gamma_{X,I}$, and the electron-phonon interaction strength, $\Gamma_{X,ph}$, for $MoS_2$, $WS_2$, and the $Mo_{0.6}W_{0.4}S_2$ alloy.

|  | $MoS_2$ | $WS_2$ | $Mo_{0.6}W_{0.4}S_2$ |
|---|---|---|---|
| $E_{0A}$ (eV) | 1.80± 0.01 | 1.96± 0.01 | 1.83± 0.01 |
| $E_{0B}$ (eV) | 1.97± 0.01 | 2.34± 0.01 | 2.09± 0.01 |
| $S_A$ | 1.4 ± 0.2 | 1.9 ± 0.2 | 1.5 ± 0.1 |
| $S_B$ | 1.5 ± 0.1 | 2.1 ± 0.1 | 1.6 ± 0.1 |
| $\hbar\omega_A$ (meV) | 26.4 ± 2.2 | 22.8 ± 3.1 | 24.4 ± 2.5 |
| $\hbar\omega_B$ (meV) | 26.4 ± 1.9 | 16.4± 3.5 | 24.4 ± 1.5 |
| $\Gamma_{A,I}$ (meV) | 132.1 ± 0.4 | 271.8 ± 0.9 | 176.4 ± 0.1 |
| $\Gamma_{B,I}$ (meV) | 280.2 ±0.7 | 366.2 ± 0.6 | 327.8± 0.1 |
| $\Gamma_{A,ph}$ (meV) | 50.8 ± 1.8 | 72.4 ± 3.0 | 54.6 ± 1.9 |
| $\Gamma_{B,ph}$ (meV) | 60.8 ± 3.1 | 73.9 ± 1.6 | 59.7 ± 2.5 |

**3.3. TDDFT calculations of the optical absorption spectrum.** The real and imaginary parts of the dielectric functions obtained from the TDDFT calculations are shown in Figures S3 -S5 and these were used to calculate the optical absorption coefficient, $\alpha$, according to Equation S3. The calculated absorption coefficients for $MoS_2$ and $WS_2$ are shown in Figure 4, together with the experimental data at 293 K. The optical absorption coefficients were obtained as described in Section 2.2, using the spectra of *T* and *R* in Figure S1. The calculations reproduce the relative energies of the A and B excitons very well, see also Table 2. In addition, the calculations reproduce the magnitude of the optical absorption coefficient to within a factor two.

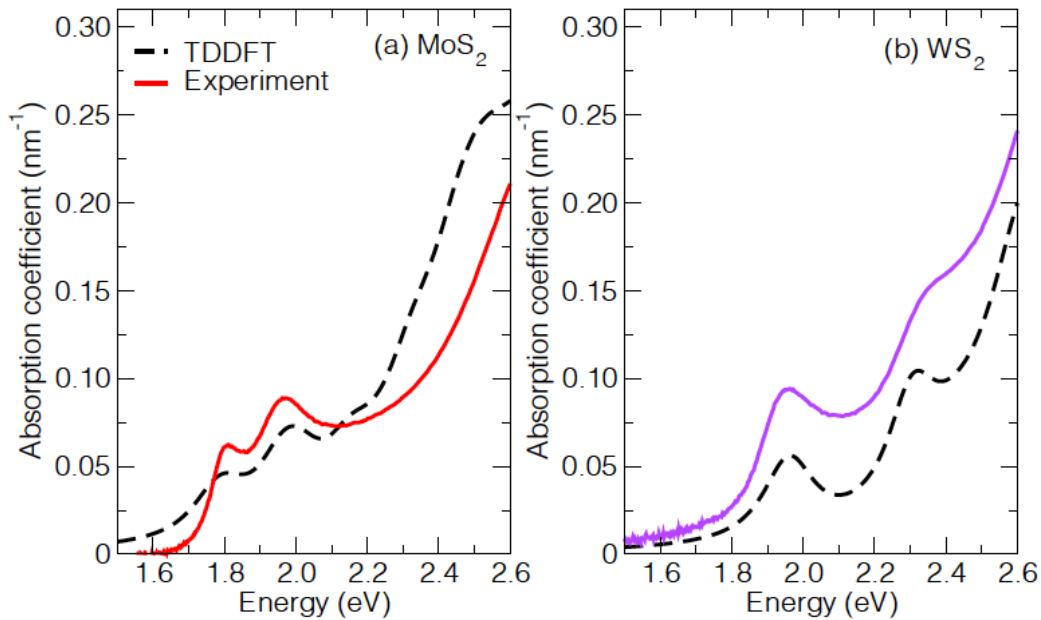

**Figure 4.** Absorption coefficient, $\alpha$, obtained from TDDFT calculations (black dashed curves) together with the experimental results at 293 K for (a) $MoS_2$ and (b) $WS_2$.

To replicate the $Mo_{0.6}W_{0.4}S_2$ alloy, we used a periodic crystal structure with a 2x2x1 supercell resulting in the $Mo_{0.625}W_{0.375}S_2$ alloy, see Figure 5. One unit cell then contains 5 Mo atoms, 3 W atoms and 16 S atoms that are arranged in two layers bonded by van der Waals forces. By permutation of the 5 Mo and 3 W atoms one can realize 28 different arrangements. These can be categorized into two groups: 1) 4 'heterogeneous' structures in which every other layer contains only Mo atoms and the layers in between contain also W atoms, and b) 24 'homogeneous' structures in which both layers contain Mo and W atoms. Applying the symmetry operations of translation, rotation, mirror planes, and their combinations, we obtain three physically distinct structures (I, II, and III), as shown in Figure 5.

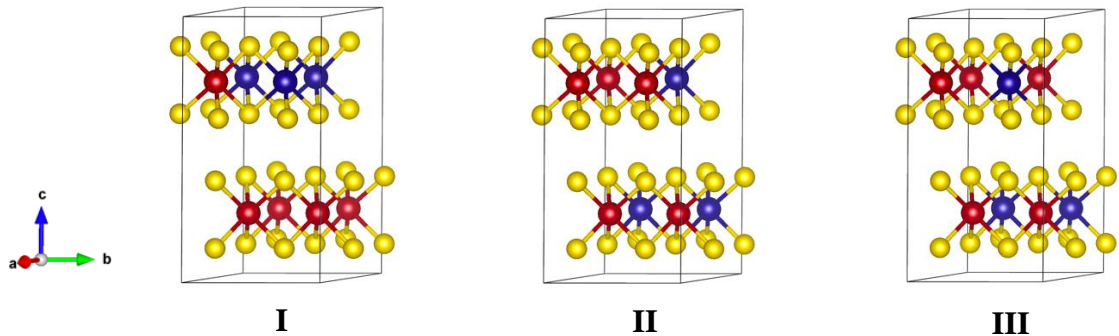

**Figure 5**. The three physically distinct arrangements of atoms in the $Mo_{0.625}W_{0.375}S_2$ alloy. Each 2x2x1 supercell of the $Mo_{0.625}W_{0.375}S_2$ alloy shows the different arrangements of metal and chalcogen atoms where the Mo atoms are red, the W atoms are blue and the S atoms are yellow. The heterogeneous structure I has alternating layers of Mo atoms only and layers containing both Mo and W atoms. In the homogenous structures II and III all layers contain Mo and W atoms.

The calculated optical absorption coefficient of the $Mo_{0.625}W_{0.375}S_2$ alloy with heterogenous structure I is shown in Figure 6(a), together with the experimental spectrum. The presence of three peaks in the calculated spectrum disagrees with the two excitonic peaks in the experimental spectrum. The third peak calculated for structure I could be due to excitons having some charge transfer character. For such excitons the electron would have a somewhat larger probability to reside on Mo atoms, while the hole is preferentially present on W atoms. Interestingly, the calculated spectra of structures II and III in Figure 6(b) agree with the experimental spectrum. The relative energies of the A and B exciton, as well as the magnitude of the optical absorption coefficient are very well reproduced by these structures, see also Table 2. From this we infer that the Mo and W atoms in the ALD-grown films are to a large extent mixed homogenously, as in structures II and III. This agrees with the previously reported random arrangement of Mo and W atoms in alloys grown by chemical vapor transport.[14] To investigate the spatial distribution of the electron and the hole within an exciton in structure I, calculations at a higher level of theory than TDDFT are needed; e.g. by describing excitons on the basis of the Bethe-Salpeter equation.[53]

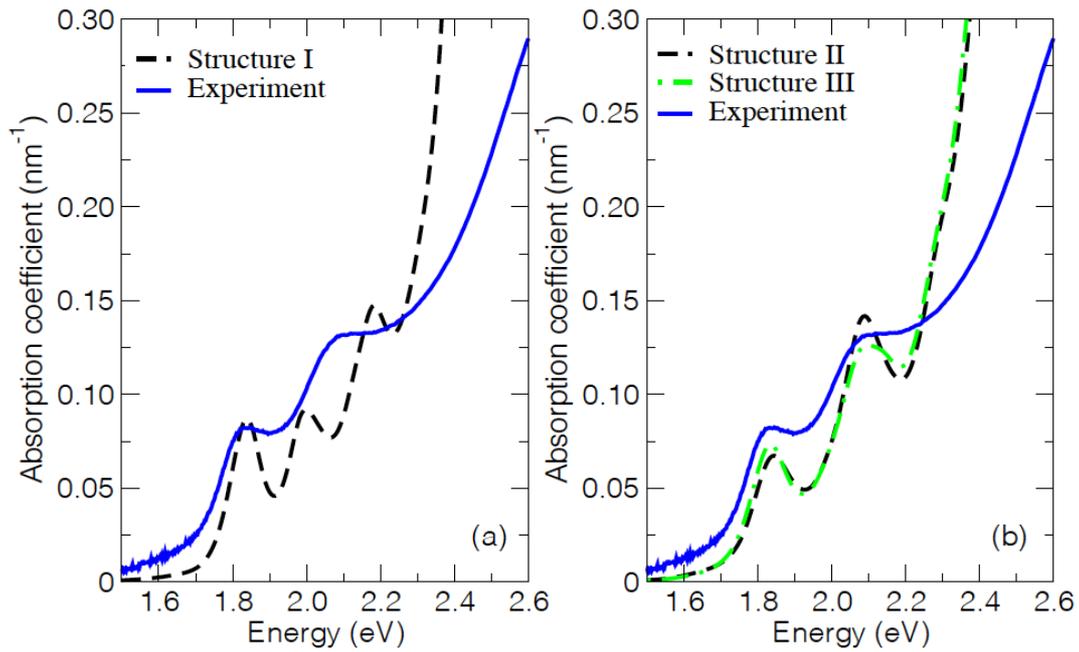

**Figure 6.** The calculated absorption coefficient of (a) structure I and (b) structures II and III of the $Mo_{0.625}W_{0.375}S_2$ alloy, together with the experimental spectrum at room temperature (293 K).

**Table 2.** Energies of the A and B excitons in $MoS_2$, $WS_2$ and the $Mo_{0.6}W_{0.4}S_2$ alloy. The last two columns show the energy difference between the exciton energies from experiments (exp.) and the TDDFT calculations.

|  | $MoS_2$ | $WS_2$ | $Mo_{0.6}W_{0.4}S_2$ | $Mo_{0.625}W_{0.375}S_2$ |
|---|---|---|---|---|
| $E_A$ (exp.) eV | 1.80 | 1.96 | 1.83 | |
| $E_B$ (exp.) eV | 1.97 | 2.34 | 2.09 | |
| $E_B - E_A$ (exp.) meV | 170 | 380 | 260 | |
| $E_B - E_A$ (TDDFT calc.) meV | 179 | 353 | | 255 |

## 4. CONCLUSION

We performed a combined experimental and time-dependent density functional theory (TDDFT) study of the optical absorption/transmission spectra of ALD-grown thin films of $MoS_2$, $WS_2$ and the alloy $Mo_{0.6}W_{0.4}S_2$. The temperature dependence of the peak energies and linewidths of the A and B excitons in the alloy is close to that for $MoS_2$. This suggests that the exciton wave functions have a larger amplitude on Mo atoms than on W atoms. From comparison of the measured optical absorption spectra with those from TDDFT calculations we infer that Mo and W atoms are homogeneously distributed throughout the alloy. These results provide clear support towards structural engineering of two-dimensional van der Waals materials through atomic arrangements, extending the already rich variety of properties in this class of materials.

## ASSOCIATED CONTENT

**Supporting information**

Absorption spectra, fits of temperature-dependent optical transmission spectra of $MoS_2$, $WS_2$, and the $Mo_{0.6}W_{0.4}S_2$ alloy.

## AUTHOR INFORMATION

**Corresponding Author**

**Laurens D. A. Siebbeles** – *Optoelectronic Materials Section, Department of Chemical Engineering, Delft University of Technology 2629 HZ Delft, The Netherlands;* [orcid.org/0000-0002-4812-7495](orcid.org/0000-0002-4812-7495); Email: [L.D.A.Siebbeles@tudelft.nl](L.D.A.Siebbeles@tudelft.nl)

## ACKNOWLEDGMENTS

N.S., D.P. and L.D.A.S. acknowledge Dr. Peter Elliott and Dr. Sangeeta Sharma for a stimulating and fruitful discussion about TDDFT calculations. This research received funding from the Netherlands Organization for Scientific Research (NWO) in the framework of the Materials for sustainability and the Ministry of Economic Affairs in the framework of the PPP allowance and is also part of the NWO research program TOP-ECHO with project number 715.016.002 as well as the NWO Gravitation program "Research Centre for Integrated Nanophotonics". This work was carried out on the Dutch national e-infrastructure with the support of SURF Cooperative.

# Effects of structure and temperature on the nature of excitons in the Mo$_{0.6}$W$_{0.4}$S$_2$ alloys

**AUTHORS**

Deepika Poonia,[†] Nisha Singh,[†] Jeff J.P.M. Schulpen,[‡] Marco van der Laan,[¶] Sourav Maiti,[†] Michele Failla,[†] Sachin Kinge,[§] Ageeth A. Bol,[‡] Peter Schall,[¶] and Laurens D.A. Siebbeles*[†]

**AFFILIATIONS**

[†]*Optoelectronic Materials Section, Department of Chemical Engineering, Delft University of Technology, 2629 HZ Delft, The Netherlands*

[‡] *Department of Applied Physics, Eindhoven University of Technology, P.O. Box 513, 5600 MB Eindhoven, The Netherlands*

[¶] *Institute of Physics, University of Amsterdam, 1098 XH Amsterdam, The Netherlands*

[§]*Materials Research & Development, Toyota Motor Europe, B1930 Zaventem, Belgium*

## 1. Transmission spectra of MoS$_2$, WS$_2$, and the Mo$_{0.6}$W$_{0.4}$S$_2$ alloy together with the fractions reflected and absorbed light.

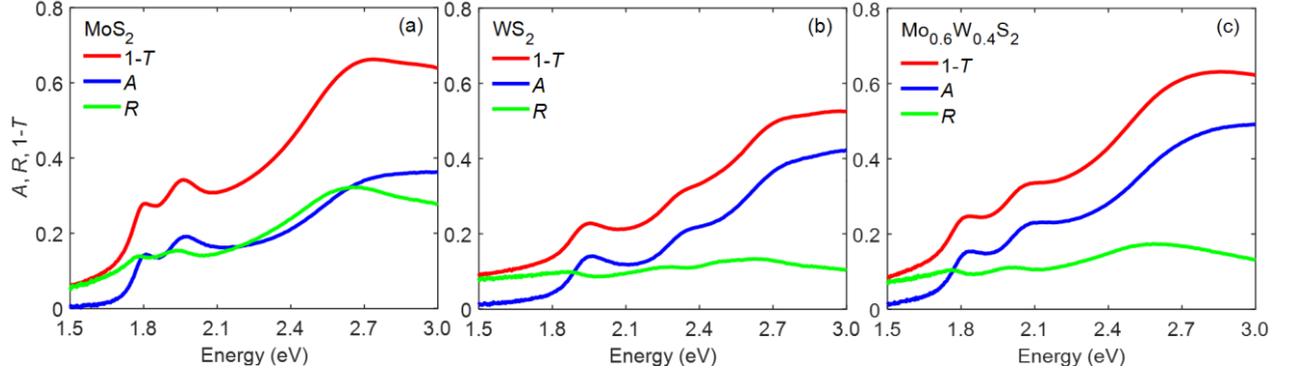

**Figure S1.** Room temperature transmission spectra (1-*T*) and fractions reflected (*R*) and absorbed (*A* = 1 - *R* - *T*) light for MoS$_2$, WS$_2$ and the Mo$_{0.6}$W$_{0.4}$S$_2$ alloy.

## 2. Fits to the temperature-dependent transmission spectra.

The optical transmission spectra were analyzed by fitting Equation S1 to the experimental results

$$1 - T(E) = \sum_{i=A,B} \frac{1}{2\pi} \frac{C_i \Gamma_i}{(E - E_i)^2 + (\Gamma_i/2)^2} + \sum_{j=1}^{2} \frac{C_j}{\sigma_j \sqrt{2\pi}} e^{\frac{-(E-E_j)^2}{2\sigma_j^2}} \qquad \text{S1}$$

with *E* the photon energy *E*. The two Lorentzian functions with amplitude $C_i$ and linewidth $\Gamma_i$ describe the A and B exciton peaks. The Gaussian functions describe optical reflection, below band gap absorption due to defects and the broad C absorption feature at higher energy. Figure S2 shows the fit results together with the experimental spectra.

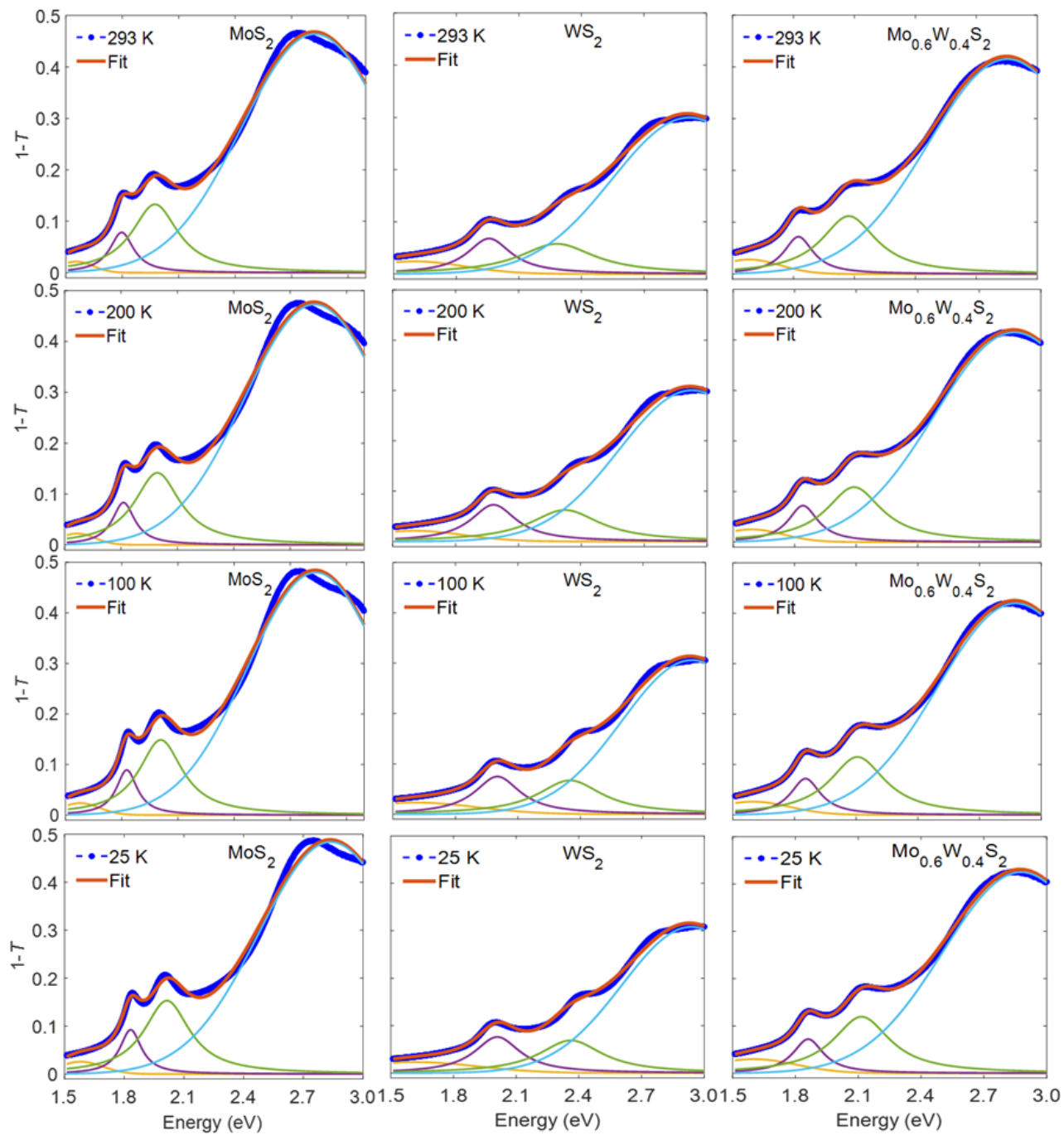

**Figure S2.** Fits to the experimental transmission spectra of MoS$_2$, WS$_2$, and the Mo$_{0.6}$W$_{0.4}$S$_2$ alloy at 293 K, 200 K, 100 K, and 25 K, respectively. The dark blue curves are the experimental spectra, the red curve is the fit of Equation S1 with the Lorentzian functions shown in purple and green and the Gaussians in yellow and light blue.

**3. Relation between absorption coefficient and dielectric function.** The optical absorption coefficient, $\alpha$, can be obtained from the real ($\varepsilon_1$) and imaginary ($\varepsilon_2$) parts of the dielectric function using the following relations[1].

$$n = \sqrt{\frac{1}{2}\left[[\varepsilon_1^2 + \varepsilon_2^2]^{1/2} + \varepsilon_1\right]}, \qquad k = \sqrt{\frac{1}{2}[[\varepsilon_1^2 + \varepsilon_2^2]^{1/2} - \varepsilon_1]} \qquad \text{S2}$$

and

$$\alpha = 2\omega k/c, \qquad \text{S3}$$

where $n$ is the refractive index, $k$ the extinction coefficient and $\omega$ the radian frequency of the light, respectively.

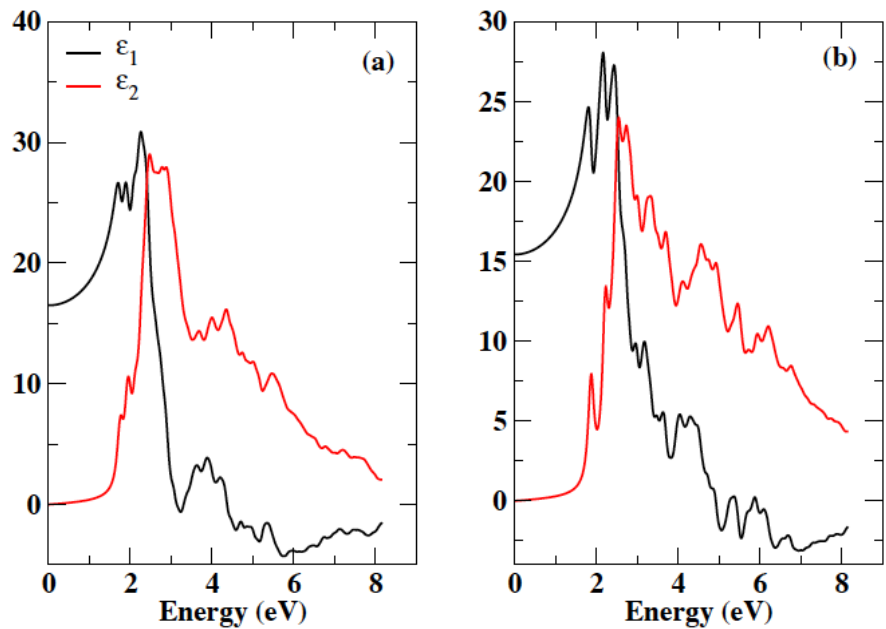

**Figure S3.** The real and imaginary part of the dielectric function obtained from the TDDFT calculations for (a) $MoS_2$ and (b) $WS_2$.

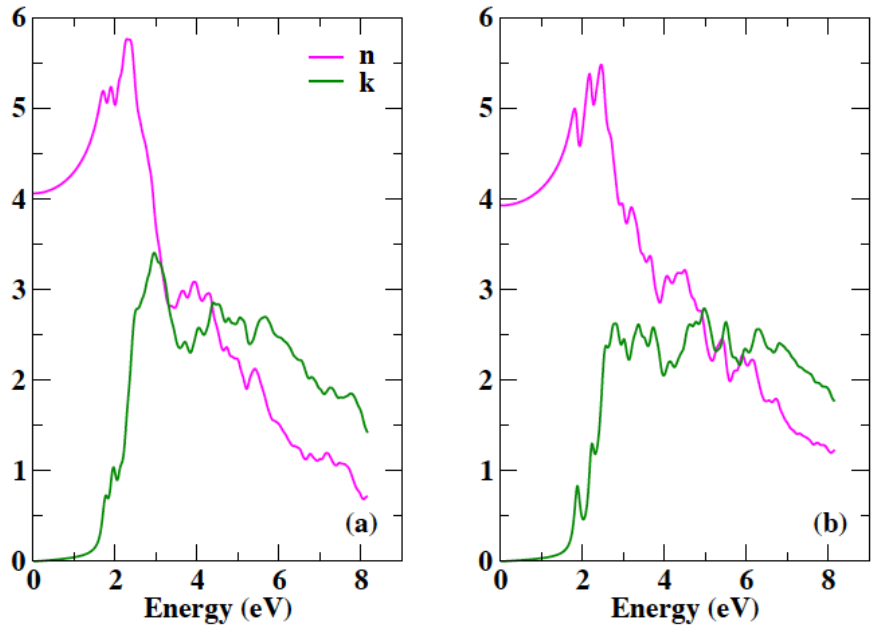

**Figure S4.** The refractive index and extinction coefficient obtained from Equation S2 for (a) $MoS_2$ and (b) $WS_2$.

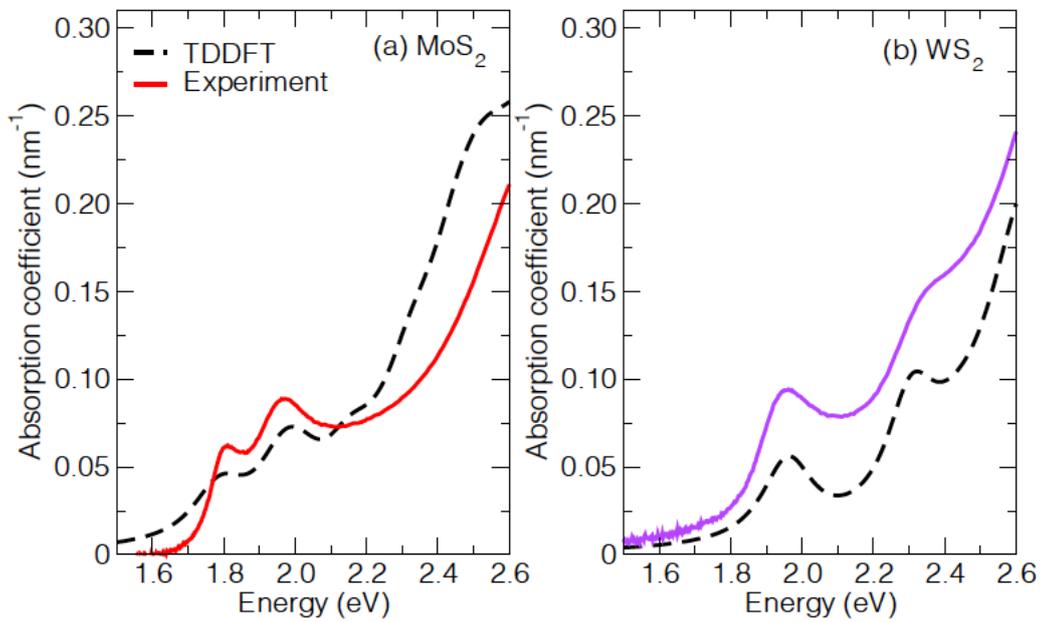

**Figure S5.** The absorption coefficient obtained from Equation S3 for (a) $MoS_2$ and (b) $WS_2$. The TDDFT results have not been shifted in energy.